\documentclass{aa}  

\usepackage{graphicx}
\usepackage{hyperref}
\usepackage{bookmark}
\usepackage{hyperref}
\usepackage{txfonts}
\usepackage[T1]{fontenc}
\usepackage{graphicx}
\usepackage{xcolor}
\usepackage{amsmath}
\usepackage{verbatim}
\usepackage{float}
\usepackage{lmodern}
\usepackage{anyfontsize}
\usepackage{comment}
\usepackage[switch]{lineno}  
\usepackage{newtxtext,newtxmath}
\usepackage{wasysym}
\usepackage{orcidlink}

\usepackage[T1]{fontenc}
\usepackage{graphicx}
\usepackage{xcolor}
\renewcommand{\textcolor}[2]{#2}
\begin{document} 
   \title{Determining the absolute chemical abundance of nitrogen and sulfur 
in the  quasar outflow of 3C298}

   \author{M. Dehghanian\inst{1}\orcidlink{0000-0002-0964-7500} 
          \and
          N. Arav\inst{1}\orcidlink{0000-0003-2991-4618}
          \and
          M. Sharma\inst{1}\orcidlink{0009-0001-5990-5790}
          \and
          D. Byun\inst{1}\orcidlink{0000-0002-3687-6552}
          \and
          G. Walker\inst{1}\orcidlink{0000-0001-6421-2449}}

   \institute{Department of Physics, Virginia Tech, Blacksburg, VA 24061, USA\\
              \email{dehghanian@vt.edu} }

   \date{Received XX, YYYY; accepted XX, YYYY}

 
  \abstract
   {Quasar outflows are key players in the feedback processes that influence the evolution of galaxies and the intergalactic medium. The chemical abundance of these outflows provides crucial insights into their origin and impact.}
   {To determine the absolute abundances of nitrogen and sulfur and the physical conditions of the outflow seen in quasar 3C298.}
   {We analyze archival spectral data from the Hubble Space Telescope (HST) for 3C298. We measure Ionic column densities from the absorption troughs and compare the results to photoionization predictions made by the Cloudy code for three different spectral energy distributions (SED), including MF87, UV-soft, and HE0238 SEDs. We also calculate the ionic column densities of excited and ground states of \ion{N}{iii} to estimate the electron number density and location of the outflow using the Chianti atomic database. }
   {The MF87, UV-soft, and HE0238 SEDs yield nitrogen and sulfur abundances at super-solar, solar, and sub-solar values, respectively, with a spread of 0.4-3 times solar. Additionally, we determined an electron number density of $\text{log}(n_{e})\geq 3.3$ cm$^{-3}$, with the outflow possibly extending up to a maximum distance of 2.8 kpc.}
  {
  Our results indicate solar metallicity within a 60$\%$ uncertainty range, driven by variations in the chosen SED and photoionization models. This study underscores the importance of SED's impact on determining chemical abundances in quasars' outflows. These findings highlight the necessity of considering a wider range of possible abundances, spanning from sub-solar to super-solar values.
}


   \maketitle
%

\section{Introduction}
Quasar outflows are essential components of feedback mechanisms that regulate the evolution of galaxies and the intergalactic medium (e.g. \citet{silk98, scan04}). These powerful winds can impact star formation within their host galaxies and contribute to metal enrichment in the surrounding environment.
Outflows are frequently observed in quasar spectra as blueshifted absorption troughs. These absorption outflows are invoked as potential contributors
to AGN feedback processes (e.g. \citet{silk98, scan04,yuan18,vayner21, he22}). 


Absorption lines in rest-frame UV spectra are generally classified into three categories based on their width: broad absorption lines (BALs) with widths of $\geq$2000 kms$^{-1}$, narrow absorption lines (NALs) with widths of $\leq$500 kms$^{-1}$, and an intermediate group referred to as mini-BALs \citep{itoh20}. 
While broad absorption line (BAL) outflows have been extensively studied \citep{vest03,gabel06}, less is known about narrow absorption line (NAL) outflows. 

We can determine the chemical abundances of an outflow by comparing the ionic column densities measured from the absorption lines across the spectrum with the results of photoionization analyses \citep{gabel06, arav07}. 
The primary advantage of using absorption lines over emission lines in abundance studies lies in their ability to provide diagnostics that are less dependent on the temperature and density (\cite{ham98, borg12}). Since BALs present challenges for determining abundances due to their complex nature, with lines often being blended and heavily saturated, we use NALs to determine the abundances. This study focuses on a NAL outflow in the quasar 3C298, where we use archival HST/FOS data to determine the absolute abundances of nitrogen and sulfur, along with physical conditions such as electron number density.

While not many studies have focused on the abundance of quasar's outflows, a majority of the available literature reports super-solar abundances. For instance, \cite{field05} and \cite{arav07} indicate that the outflow in their studied quasars is significantly enriched, with abundances often exceeding solar values \citep{gr10} by factors of several. \cite{gabel06} find that the absorption outflow in QSO J2233-606 has super-solar abundances of carbon, nitrogen, and oxygen by roughly a factor of 10 (see Table~\ref{tab2}). The nitrogen abundance is the most enhanced. As investigated by \cite{arav20}, the absorption outflow in NGC 7469 also exhibits super-solar abundance, with carbon and nitrogen abundances being roughly twice and four times the solar values, respectively. 
However, our analysis of the outflow seen in quasar 3C298 indicates that the abundances of nitrogen and sulfur are 0.4-3 times their solar values, while they can extend to sub and super-solar values when considering different spectral energy distributions (see Section~3.2). 
\textcolor{magenta}{These findings are consistent with recent studies, suggesting that radio-loud quasars tend to exhibit slightly sub-solar metallicity in their line-emitting regions, while radio-quiet quasars may show abundances closer to solar values. For example, the study by \cite{pun18} examines the metallicity of the broad-line region (BLR) in NGC 1275, a nearby radio-loud AGN with a powerful jet. The findings indicate that it has metallicity levels near solar, which is consistent with Seyfert galaxies and supports a scenario of moderate enrichment likely driven by jet and accretion activity. \cite{marz23} reveals that radio-quiet quasars generally have slightly sub-solar or solar metallicities, with no significant supersolar enrichment. In contrast, radio-loud quasars tend to exhibit distinctly sub-solar metallicities, often lower than their radio-quiet counterparts. In a more recent study, \cite{flo24}  focus on the BLR of AGNs across different populations in the quasar main sequence. The study reveals a systematic metallicity gradient along the quasar main sequence, from sub-solar levels in Population B quasars (specifically radio-loud ones) to extremely high, super-solar values in extreme Population A (xA) sources. \cite{vil24} observed sub-solar metallicity in the giant nebula associated with the Teacup quasar. However, within the AGN-driven bubble, their study identified solar or slightly super-solar metallicity. This highlights the complexity of metallicity distribution in AGN environments, with localized metal enrichment occurring within outflow structures. The use of multiple SEDs (MF87, UV-soft, and HE0238) allows us to assess the impact of varying radiation fields on the derived chemical abundances and constrain the conditions of the outflow in 3C298.}


It is worth mentioning \citet{pun22} conducted a detailed analysis of 3C298’s radio lobes, estimating one of the largest known long-term jet powers at $1.28 \times 10^{47}$ erg~s$^{-1}$ and identifying a weak ionizing continuum. In contrast, our study focuses on the chemical abundances of nitrogen and sulfur in the narrow absorption line outflow, offering insights into the physical conditions of the NAL outflow and its role in quasar feedback.

In the following sections, we discuss the observational data, the methodology used to derive abundances from absorption lines, and additional relevant analyses. The results presented here emphasize the importance of considering a broader range of abundances in quasar outflows and underscore the necessity for further research in this relatively under-explored field.

This paper is structured as follows. In Section 2, we present the observational data used in this study, including the spectral data from quasar 3C298 and the relevant absorption lines. Section 3 details the methodology employed to derive elemental abundances from these absorption lines, with a focus on the techniques used for extraction of the ionic column densities and photoionization modeling, as well as electron number density determinations. Finally, Section 4 concludes the paper with a summary of the key findings. Additionally, Appendices A and B provide detailed error calculations related to the methods used.

Here we adopted standard $\Lambda$CDM cosmology with h= 0.677, $\Omega_{m}$= 0.310, and
$\Omega_\Lambda$ = 0.690 \citep{plan20}. We used the Python astronomy
package Astropy \citep{astro13,astro18} for our
cosmological calculations, as well as Scipy \citep{virt20},
Numpy \citep{harr20}, and Pandas \citep{reba21} for most of our numerical computations. For our plotting purposes, we used Matplotlib \citep{hunt07}.
\section{Observations}
QSO 3C298 is a bright radio source, at Redshift z=1.4362 (based on the NASA/IPAC Extragalactic Database \footnote{NED:\url{https://ned.ipac.caltech.edu/}} and also \cite{diego07}),
with J2000 coordinates at RA=14:19:08.18, DEC=+06:28:34.80. Table~1 of \cite{pun22} summarizes the radio data available for this object. 
The quasar was observed using the HST Faint Object Spectrograph (FOS) on two separate occasions, on August 15th, 1996, as part
of HST \citet{ham96}(PI: Hamann). These observations have exposure times of 1.6 and 2.4 ks, and both utilize the G270H grating with a central wavelength of 2650\AA. We co-added these two spectra and binned them by three pixels to prepare the data for further analysis. 

\begin{figure*}
\includegraphics[width=7 in]{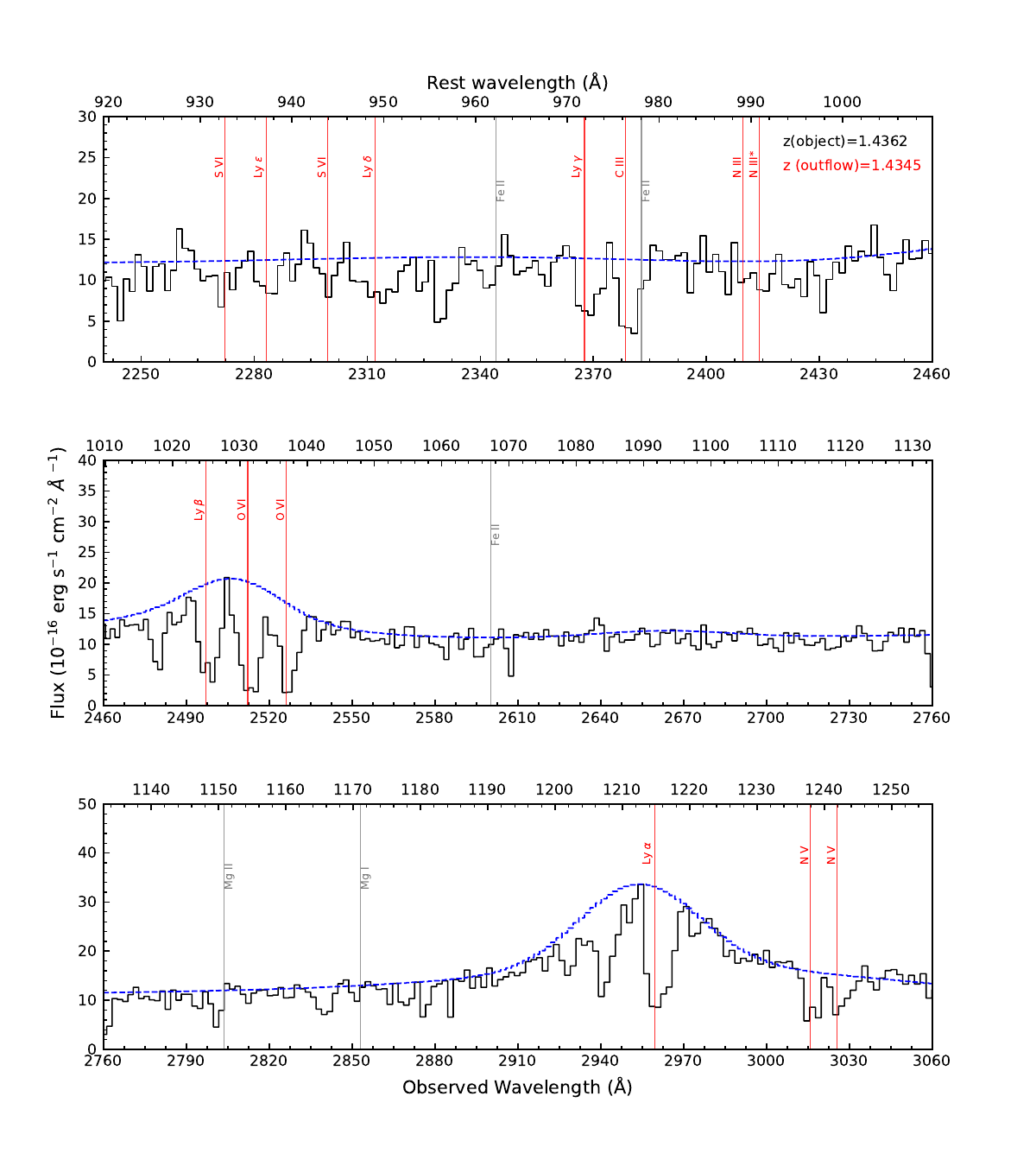}
\caption{The spectrum of 3C298 observed by the HST/FOS in August 1996. The absorption features of the outflow with a velocity of --210 km/s are marked with red lines, while the grey lines show the absorption from the interstellar medium (ISM). The dashed blue line shows our unabsorbed emission model.}
            \label{figFlux}%
\end{figure*}

After obtaining the relevant data from the Mikulski Archive for
Space Telescopes (MAST), we detected an outflow at a velocity of $v_{\textrm{centroid}}$=--210 km s$^{-1}$, with blue-shifted ionic absorption lines denoted by red vertical lines in Figure~\ref{figFlux}. 
\section{Analysis}
The identified absorption lines in the spectra are shown in Figure~\ref{figFlux}. We exclude the \ion{C}{iii} $\lambda$977.02\AA\  and \ion{O}{vi} doublet absorption lines ($\lambda\lambda$ 1031.91,1037.61\AA\  ) from our analysis since they are observed to be very deep (especially compared to the Ly$\alpha$ absorption line), suggesting that these absorption lines are most likely saturated. Also, among the available Lyman lines, we focus on using Ly$\epsilon$, as it is the least saturated of the series.
\subsection{Ionic column density determinations}
Determining the ionic column densities ($N_{\textrm{ion}}$) of the absorption lines is the first step in understanding the physical characteristics of the outflow. The "apparent optical depth" (AOD) is one method to measure the ionic column densities using the absorption lines. This approach assumes that the outflow uniformly and completely covers the source\citep{spit68,sava91}. 
When using the AOD method, the column density can be determined using Equations~\ref{eq-1} and \ref{eq0} below \citep{spit68, sava91,arav01}:
\begin{equation}
I(v)=e^{-\tau(v)}, \label{eq-1}
\end{equation}
\begin{equation}
N_{\textrm{ion}}=\frac{3.77\times 10^{14}}{\lambda_{0}f}\times \int \tau(v)~dv 
 [\textrm{cm}^{-2}],
\label{eq0}
\end{equation}
\noindent where:
   \[
      \begin{array}{lp{0.8\linewidth}}
         I(v)  &  normalized intensity profile as a function of velocity    \\
         \tau(v)             & optical depth of the absorption trough             \\
         \lambda_{0}          & transition's wavelength     \\
         f             & oscillator strength of the transition \\
      \end{array}
   \]

To determine the ionic column densities of the Ly$\epsilon$, \ion{N}{iii}~$\lambda989.80$\AA\, and \ion{N}{iii*}~$\lambda$991.58\AA\, we opted for the AOD technique, given that these ions are all singlets. Note that the absorption troughs of these mentioned ions are much shallower than other absorption lines (particularly compared to the Ly$\alpha$), ensuring that they are not saturated and their column density measurements via the AOD method are reliable. For the case of \ion{S}{vi}, we only consider its red trough (\ion{S}{vi}~$\lambda$944.52\AA) and treat it as a singlet. The reason is that the red trough exhibits lower noise levels compared to the blue trough, making its measurements more reliable.  For more details regarding the AOD method, see \cite{arav01}, \cite{gabel03}, and \cite{byun22c}. 

\begin{figure*}
\centering
\includegraphics[width=7 in]{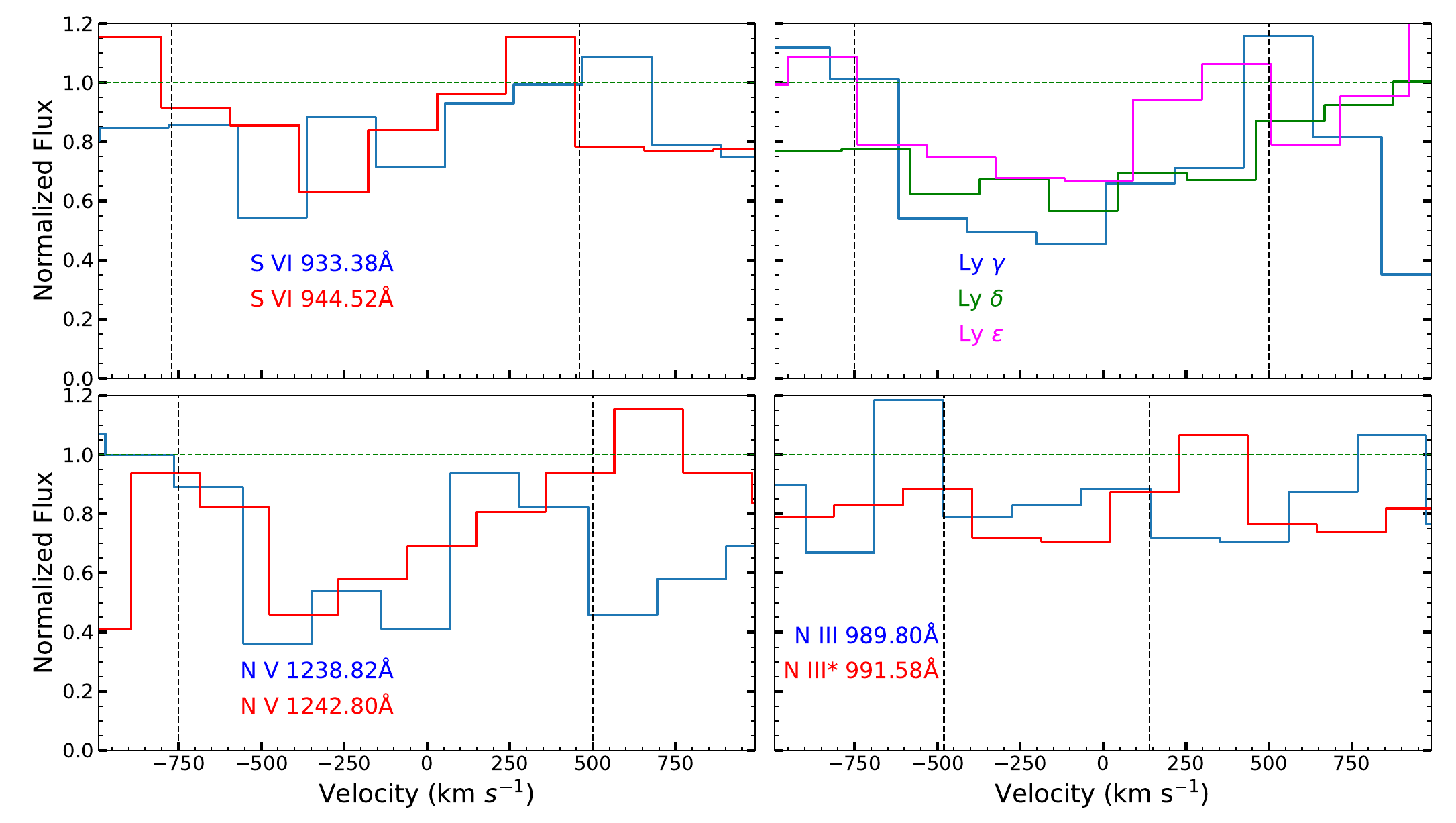}
\caption{Normalized flux versus velocity for blue-shifted absorption lines detected in the spectrum of 3C298. The horizontal green dashed horizontal line shows the continuum level,
and the vertical black dashed lines show the integration region for each line.}
            \label{figvelo}%
\end{figure*}

The second method is the partial covering (PC) method, which will be applied when there are two or more lines originating from the same lower level. This approach assumes a homogeneous source, partially covered by the absorption outflow \citep{barl97, arav99a, arav99b}. By using the PC method and deducing a velocity-dependent covering factor, we ensure that phenomena such as non-black saturation are considered \citep{kool02}. In this method the covering fraction $C(v)$ and the optical depth $\tau(v)$ are calculated using  Equations~\ref{eq3}$\&$~\ref{eq4}\citep{arav05}. For a doublet transition where the blue component has twice the oscillator strength as the red component: 
\begin{eqnarray}
    I_R(v)-[1-C(v)]=C(v)e^{-\tau(v)} \label{eq3}
\end{eqnarray}
and
\begin{eqnarray}
    I_B(v)-[1-C(v)]=C(v)e^{-2\tau(v)}\label{eq4}
\end{eqnarray}
\noindent in which $I_R(v)$ and $I_B(v)$ are the normalized intensity of the red and blue absorption feature (\textcolor{magenta}{as shown in Figure\ref{figvelo}}), respectively, while $\tau(v)$ is the optical depth profile of the red component. 

\begin{figure}
\includegraphics[width=\columnwidth]{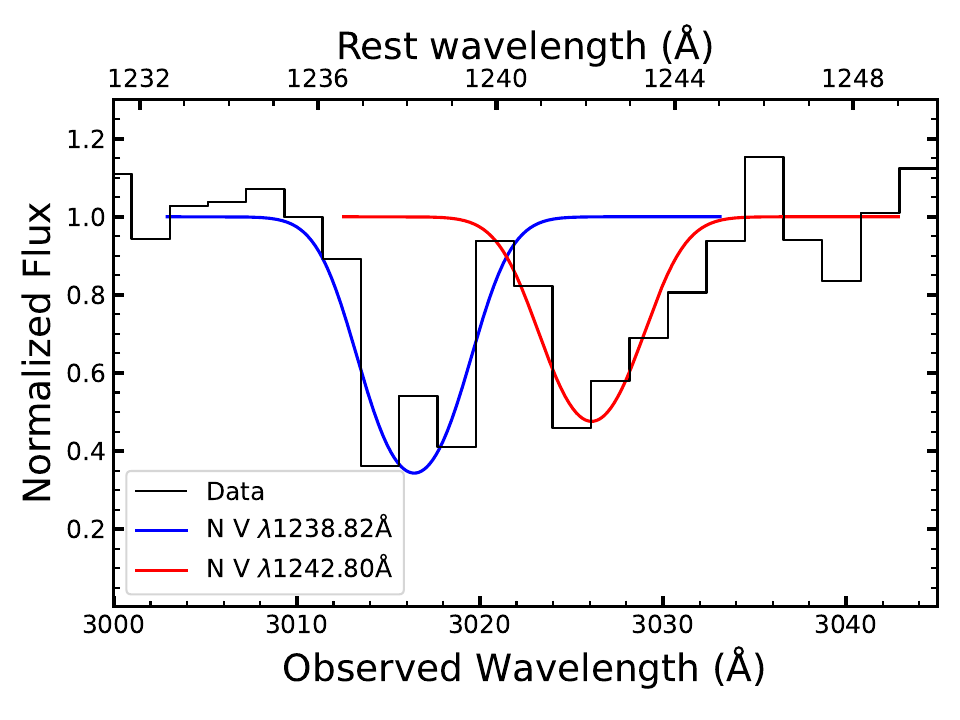}
      \caption{Gaussian modelings of the \ion{N}{v}~$\lambda$1238.82\AA\ and \ion{N}{v}~$\lambda$1242.80\AA\  absorption trough.}
         \label{figNV}
\end{figure}

We apply the PC method to the \ion{N}{v} doublet ($\lambda\lambda$1238.82, 1242.80\AA).
Due to the moderate S/N of the data, some of the red doublet flux values are lower than the blue flux values at the same velocity (see Figure~\ref{figvelo}). Such a situation does not allow for a PC solution.  We, therefore, model each trough with a Gaussian in flux space, where both Gaussians have the  same centroid velocity ($v_{\textrm{centroid}}$) and width (see Figure~\ref{figNV}).  We then perform the PC method on the resultant Gaussians and convert the resultant $\tau(v)$ and $C(v)$ to column density.
For a detailed explanation of both methods and to better understand the logic behind them and get a sense of their 
mathematics, see \cite{barl97,arav99a,arav99b, kool02, arav05,borg12a,byun22b, byun22c}.

For each of the mentioned absorption lines, we 
used the redshift of the 
outflow (z$_{\textrm{outflow}}$=1.4345) to transfer the spectrum from 
wavelength space to velocity space (see Fig.~\ref{figvelo}). When performing column density 
calculations, we have chosen a velocity range of $-$750 to $+$500 
km~s$^{-1}$ as our integration range for \ion{S}{vi}, Ly$\epsilon$, and \ion{N}{v}. For \ion{N}{iii}, and \ion{N}{iii*} ions, the integration range is smaller, as the absorption trough is narrower (see Fig.~\ref{figvelo}). The integration ranges for each ion are indicated by vertical dashed lines. We have also included Ly$\gamma$ and Ly$\delta$ in Figure~\ref{figvelo} for comparison purposes.

\begin{figure}
\includegraphics[width=\columnwidth]{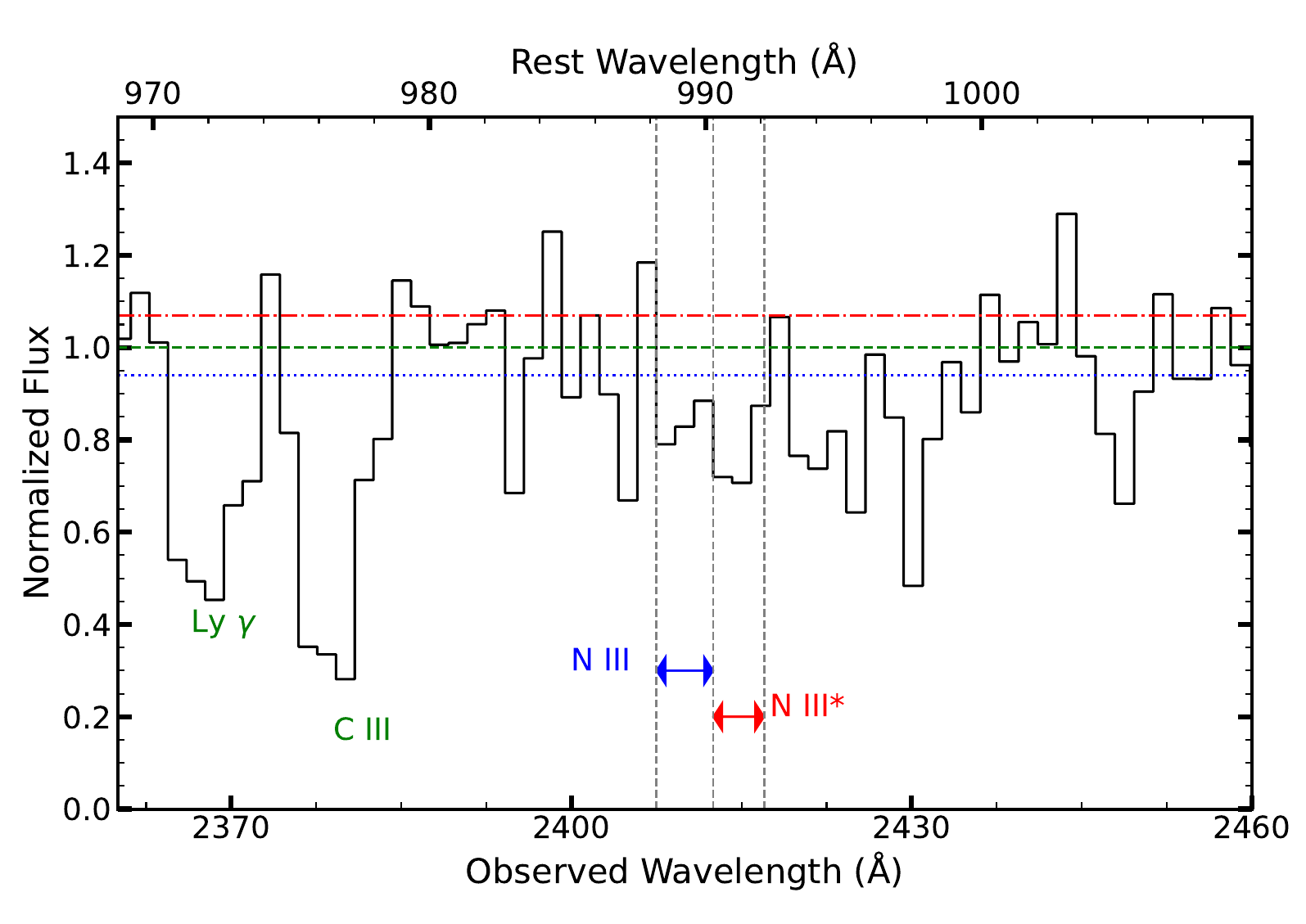}
      \caption{Absorption line profiles for the quasar 3C298, plotted against velocity in km~s$^{-1}$ in the regions around \ion{N}{iii} and \ion{N}{iii*}. The normalized flux is shown on the y-axis, with the absorption lines of Ly$\epsilon$, \ion{C}{iii}, \ion{N}{iii} and \ion{N}{iii*} labeled. The dashed green line shows the local continuum model, while the red dashed and blue dotted lines show $\pm5\%$ uncertainties. The integration range for \ion{N}{iii} and \ion{N}{iii*} is also shown with blue and red arrows, respectively.}
         \label{figCont}
\end{figure}

Before finalizing our results, it is essential to estimate the uncertainties arising from systematic errors. To achieve this, we repeated the column density calculations while adjusting for both upper and lower thresholds of the local continuum (as shown in Figure~\ref{figCont}). The final adopted errors for column densities of Ly$\epsilon$, \ion{N}{iii}, \ion{N}{iii*}, and \ion{S}{vi}  are determined by quadratically combining the AOD errors with the systematic errors arising from 5$\%$ uncertainty in the local continuum level. For \ion{N}{V}, errors are directly derived from the Gaussian fits.  Table~\ref{tab1} summarizes the results of our $ N_{\textrm{ion}}$ measurements,
along with their associated uncertainties. 

\subsection{Photoionisation solution}
\label{sec:photosol}
At this stage, we use the ionic column densities to determine the total hydrogen column density ($N_{\textrm{H}}$)
and the ionization parameter ($U_{\textrm{H}}$) following established methodologies \citep[e.g.,][]{xu19, byun22a, byun22b, byun22c, walk22,deh24}. We employ the Cloudy code \citep{cloud23} to generate a grid of ($N_{\textrm{H}}$) and ($U_{\textrm{H}}$) values, allowing us to compare theoretical predictions with observational data that are available in Table~\ref{tab1}. 
\begin{figure}
\includegraphics[width=\columnwidth]{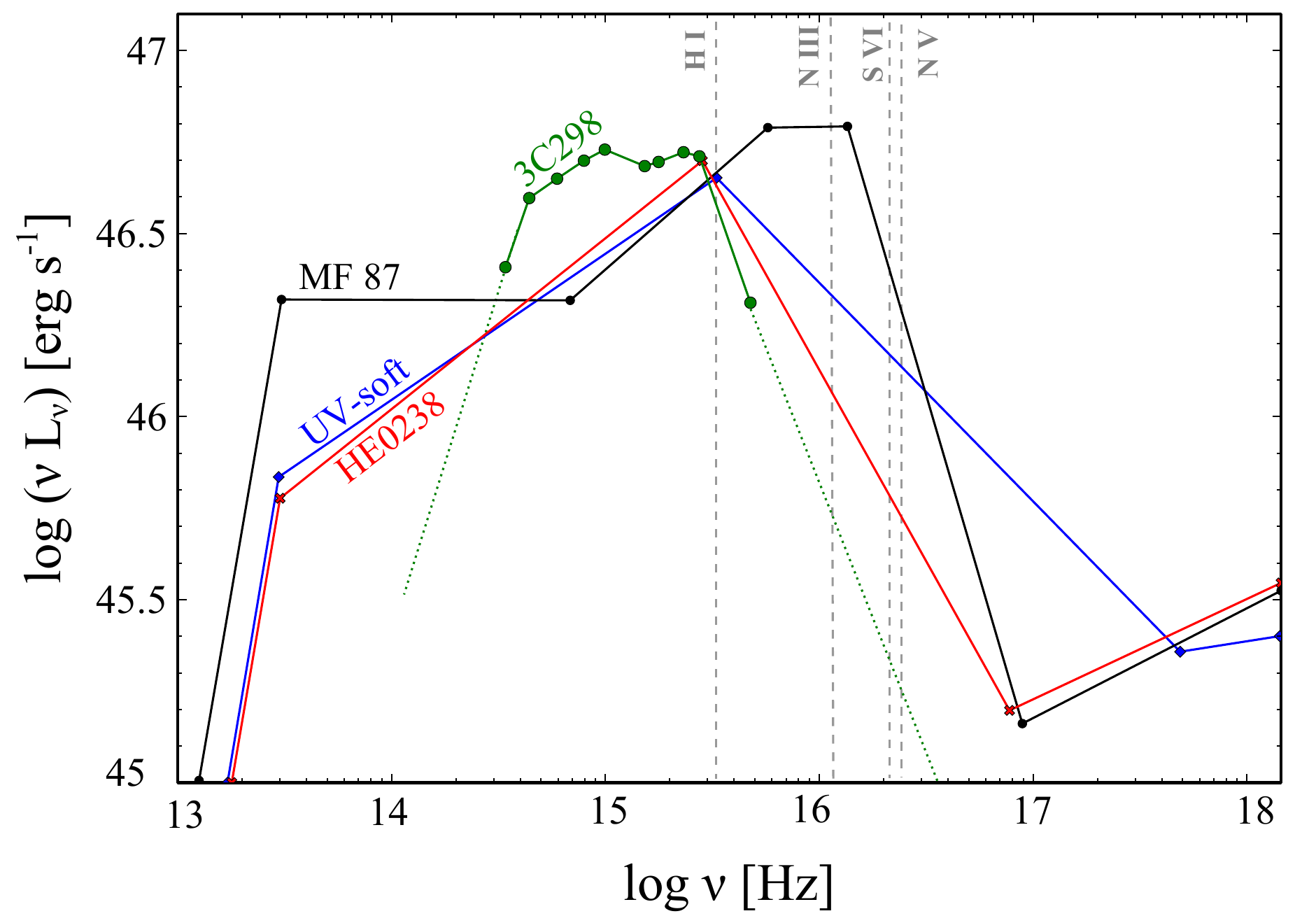}
      \caption{\textcolor{magenta}{The three SEDs used in the analysis (from \cite{arav13})along with the 3C298's SED generated by \cite{pun22}. The vertical dashed lines show the ionization energy for each labeled ion. }}
         \label{figSed}
\end{figure}
\textcolor{magenta}{The photoionization state of an outflow is influenced by the spectral energy distribution (SED) incident upon it. To evaluate how the choice of SED affects our abundance determination results, we conduct our analysis using three different SEDs: HE0238, MF87, and UV-soft, as shown in Figure~\ref{figSed}. This figure also displays the SED generated for 3C298 by \cite{pun22} (shown in green). The ionizing portion of the SED is critical for our study. As illustrated in Figure~\ref{figSed}, the 3C298 SED from \cite{pun22}' s value at log$(\nu)\approx15.7$ [Hz]
is only 0.1 dex lower than HE0238's $\nu L_{\nu}$ at the same frequency, indicating general agreement in the ionizing range. However, we exclude the 3C298 SED from our analysis due to its limited data points in this crucial region.} 
\subsubsection{HE0238 SED}
First, we adopt the SED of quasar HE0238-1904 \citep[hereafter HE0238,][]{arav13} and perform Cloudy simulations. HE0238 is the most empirically determined SED in the extreme ultraviolet (EUV) region, providing a reliable model for studying ionization processes in AGN outflows.
\begin{table}[ht!]
\caption{\label{tab1}Ionic column densities}
\centering
\begin{tabular}{lccc}
\hline\hline
Ion&$N_{\textrm{ion}}$&Upper Uncertainty&Lower Uncertainty\\
\hline
Ly$\epsilon$           &210     &65 & $-$48\\
\ion{N}{v}           &23.5   &2.2& $-$2.2\\
\ion{N}{iii} &2.6     &1.2&$-$1.3\\
\ion{N}{iii*}    &4.6      &1.4 &$-$1.4\\
\ion{S}{vi}      &3.2     &1.3&$-$1.3\\
\hline
\end{tabular}
\tablefoot{The ionic column densities of the absorption lines detected in the 3C298 outflow. All of the column density 
values are in units of 10$^{14}$ cm$^{-2}$.}
\end{table}

\begin{figure}
\includegraphics[width=\columnwidth]{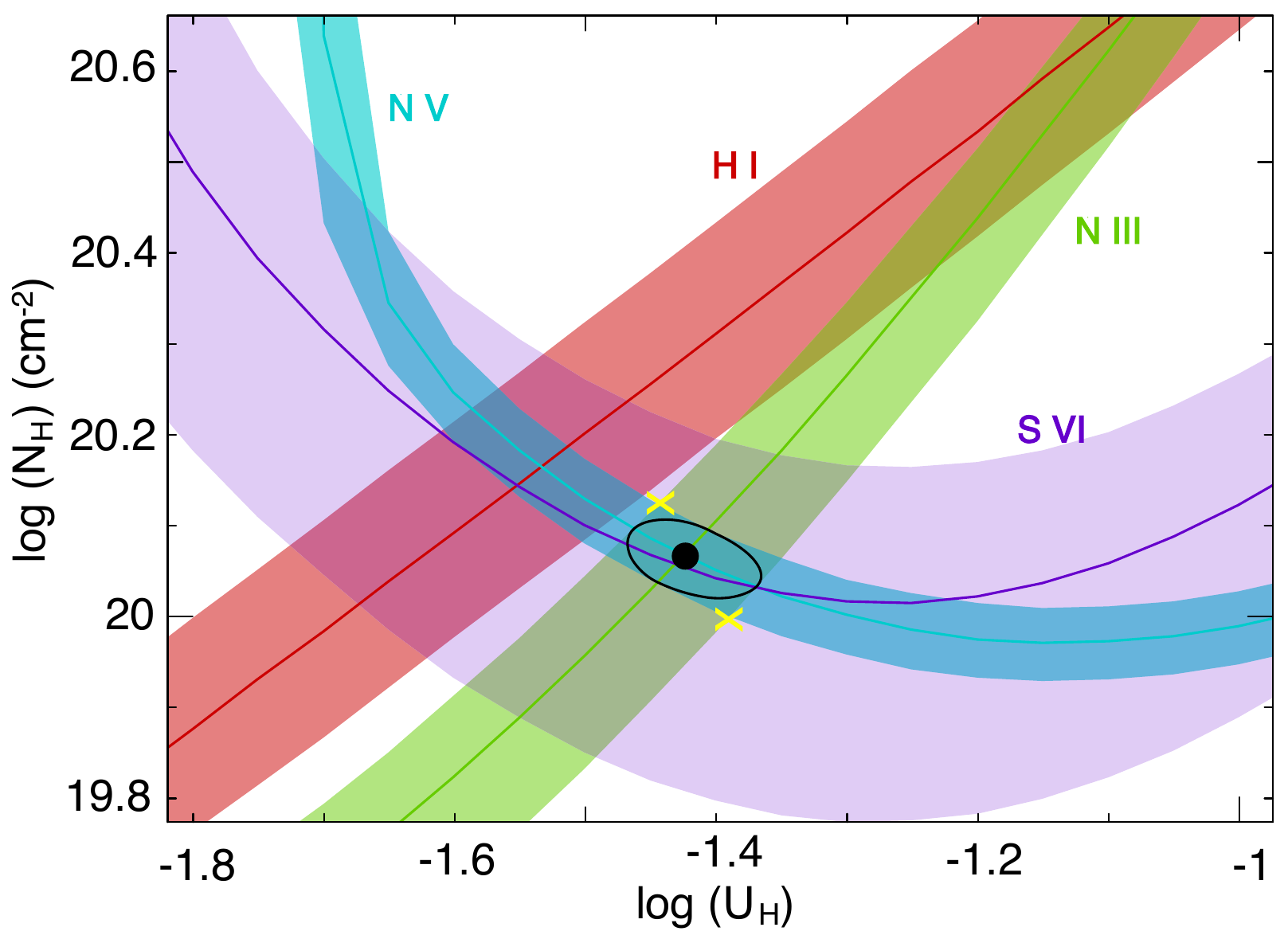}
      \caption{Photoionization solution for the outflow detected in 3C 298. Each coloured contour indicates the ionic column densities consistent with the observations (presented in Table~\ref{tab1}), assuming the HE0238 SED and solar abundances. The solid lines show the measured value, while the shaded bands are the uncertainties. The black dot shows the best solution for \ion{N}{iii} and \ion{N}{v}, with the 1$\sigma$ uncertainty shown by the black oval. The two yellow crosses show the allowed uncertainty region used to determine the absolute abundances (see Appendix\ref{app}).}
         \label{figHSN}
\end{figure}

Figure~\ref{figHSN} illustrates the results for \ion{H}{i} (estimated based on Ly$\epsilon$), \ion{N}{iii}, \ion{N}{v}, and \ion{S}{vi}, assuming solar abundances.
As this Figure shows, while there is a photoionization solution (shown by a black dot) that reproduces the \ion{N}{iii}, \ion{N}{v}, and \ion{S}{vi} ionic column densities, it does not reproduce the observed column density of \ion{H}{i} simultaneously. This discrepancy suggests that assuming solar abundances fails to account for all of the observed ionic column densities.  

The \ion{H}{i} column density predicted by the photoionization model presented in Figure~\ref{figHSN} is $1.66^{+0.61}_{-0.50}$ times lower than the observed value (see Appendix~\ref{app} for detailed calculations). This finding indicates that adjustments to the photoionization models are necessary for a more accurate model of the absorption outflow. The only free parameters in the photoionization modeling (besides the choice of SED) are the absolute abundances of nitrogen and sulfur. By decreasing the absolute abundances of nitrogen by a factor of 1.66, we can achieve a minimized-$\chi^2$ solution that reproduces the observations:
\begin{equation*}
\begin{split}
\frac{\textrm{Expected\  Abundances\ of\  N}}{\textrm{Solar Abundances of N}} = \frac{1}{1.66^{+0.61}_{-0.50}}=0.60^{+0.26}_{-0.16}\\ \rightarrow [\textrm{N}/\textrm{H}]=-0.22^{+0.16}_{-0.14}
\end{split}
\end{equation*}
Similarly, since the contour of \ion{S}{vi} nearly coincides with the crossing point of the \ion{N}{iii} and \ion{N}{v} lines, we can estimate that its abundances have to be adjusted by the same value:
\begin{equation*}
[\textrm{S}/\textrm{H}]=-0.22^{+0.16}_{-0.14}
\end{equation*}
It is worth mentioning that the notation [X/H] is commonly used to express the abundance of element X relative to the hydrogen (H), compared to the same abundance ratio in the Sun. This value is usually reported on a logarithmic scale.

\begin{figure}
\includegraphics[width=\columnwidth]{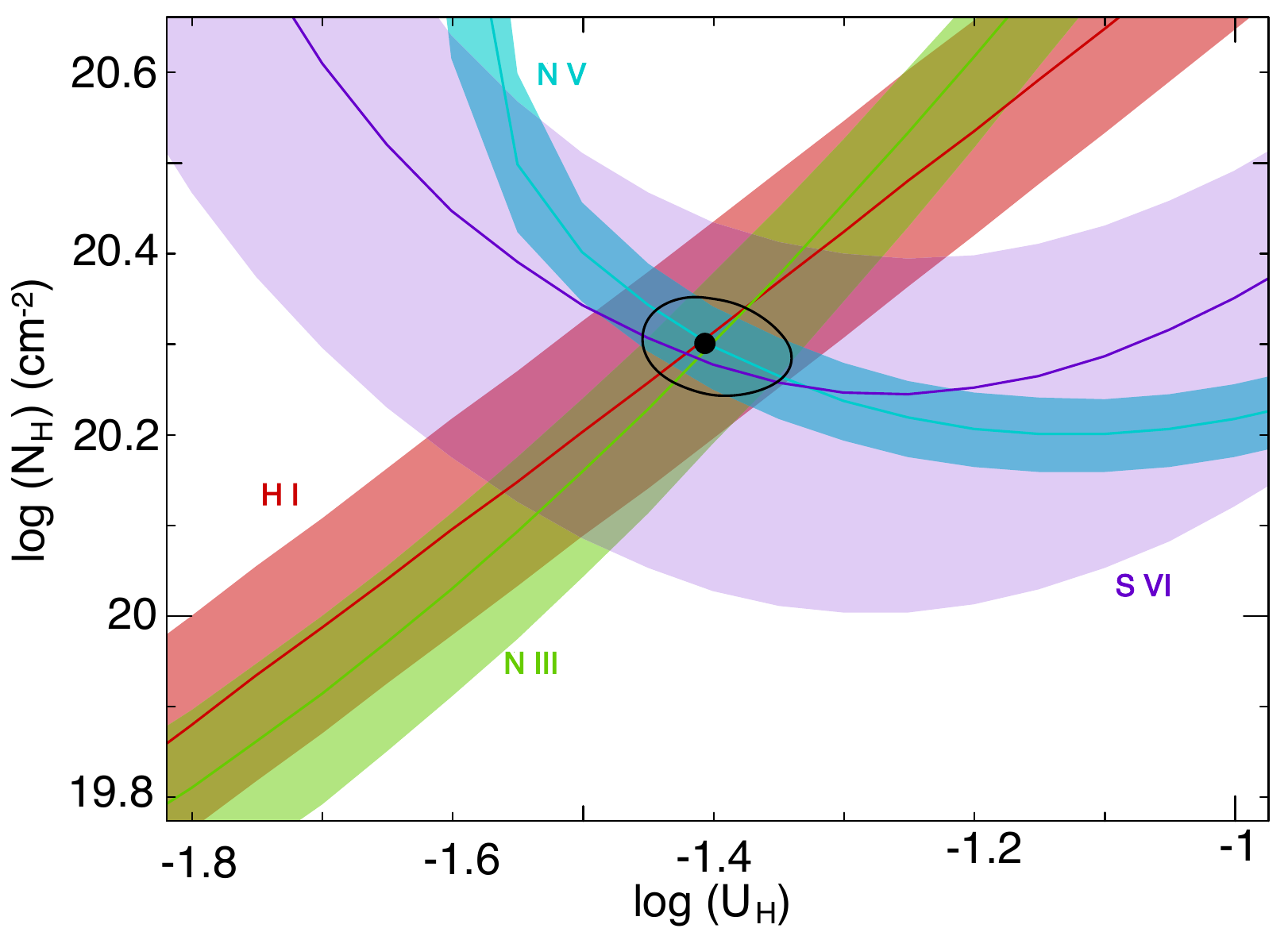}
      \caption{Photoionization solution for the outflow in 3C298, assuming the HE0238 SED and solar abundance for all elements, except for the nitrogen and sulfur. The abundances of nitrogen and sulfur are set to be 60$\%$ of their solar value with respect to hydrogen. }
         \label{figHNS-shifted}
\end{figure}

To confirm that the nitrogen and sulfur abundances must be sub-solar, we conducted a new set of Cloudy simulations, assuming solar abundances for all elements except nitrogen and sulfur, for which we scaled its abundance to 60$\%$ of their solar values (as calculated above). The results are presented in Figure~\ref{figHNS-shifted}, which displays the final photoionization solution for the outflow, assuming the HE0238 SED and adjusted abundances for nitrogen and sulfur. 
As shown in this Figure, for the HE0238 SED, assuming a sub-solar nitrogen and sulfur abundance yields a photoionization solution with
$\chi^{2}_{\textrm{reduced}}=0.01$, indicating an excellent fit for the outflow. Below, we repeat these analyses for two more SEDs.

\subsubsection{M87 SED}
\begin{figure}
\includegraphics[width=\columnwidth]{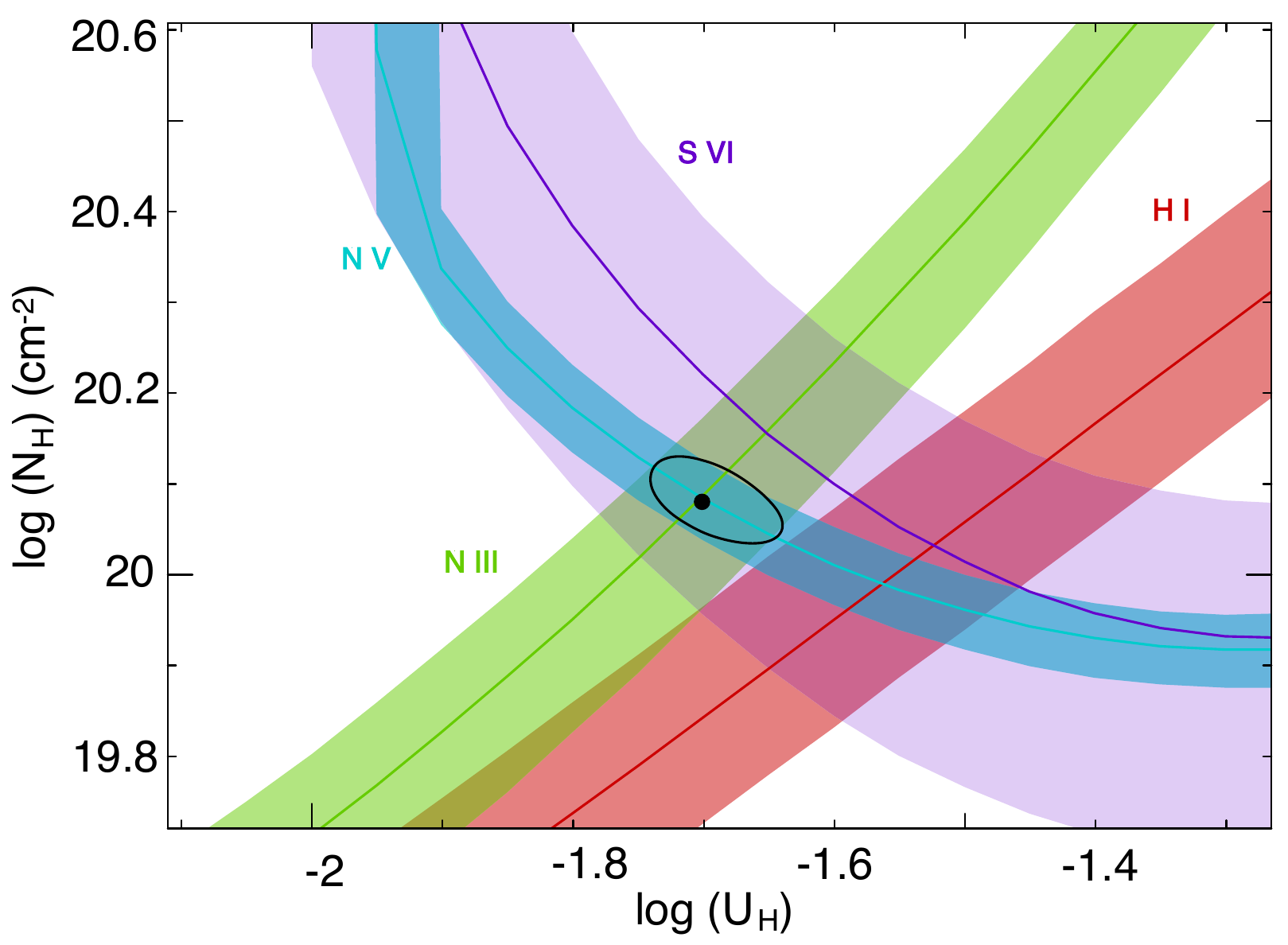}
      \caption{Same is Figure~\ref{figHSN}, utilizing the MF87 SED and solar abundances. }
         \label{figMF87}
\end{figure}

The second SED we employ is the MF87 SED, which was designed by \cite{mat87} to model the energy distribution from UV-FUV bump to the X-ray regime, and has been widely used to model radio-loud quasars.
Figure~\ref{figMF87} shows the results of the photoionization modeling using this SED and assuming solar abundances. 
As this Figure illustrates and in contrast with our previous modeling, the minimized-$\chi^{2}$ photoionization solution that reproduces the observed value of \ion{N}{iii} and \ion{N}{V} (black dot in Figure~\ref{figMF87}) overproduces hydrogen. This solution does not produce the observed column density of \ion{S}{vi} either. However, its predictions of \ion{S}{vi} are still within the measured uncertainties.

The \ion{H}{i} column density predicted by the photoionization solution
presented by the black dot in Figure~\ref{figMF87} is approximately $0.6^{+0.22}_{-0.18}$ times the observed value (error calculation follows the procedure explained in Appendix~\ref{app}). This means to have a solution that produces the observed values of nitrogen and hydrogen at the same time, we have to increase the nitrogen abundance (with respect to the hydrogen):
\begin{equation*}
\begin{split}
\frac{\textrm{Expected\  Abundances\ of\  N}}{\textrm{Solar Abundances of N}}(\textrm{with respect to H}) &= \frac{1}{0.6^{+0.22}_{-0.18}}\\
=1.66^{+0.72}_{-0.44}
\rightarrow [\textrm{N}/\textrm{H}]=0.22^{+0.16}_{-0.13}
\end{split}
\end{equation*}
We must also adjust sulfur abundances to reproduce the observations simultaneously. In Figure~\ref{figMF87}, to match the \ion{S}{vi} contour with the crossing point of \ion{N}{iii} and \ion{N}{v}: 
\begin{equation*}
\begin{split}
\frac{\textrm{Expected\  Abundance\ of\  S}}{\textrm{Solar Abundance of S}} (\textrm{with respect to N}) = 1.3^{+0.5}_{-0.5}
\end{split}
\end{equation*}
\noindent resulting in:
\begin{equation*}
\begin{split}
\frac{\textrm{Expected\  Abundance\ of\  S}}{\textrm{Solar Abundance of S}} (\textrm{with respect to H})
=\\
1.3^{+0.5}_{-0.5} \times 1.66^{+0.72}_{-0.44}=2.16^{+1.2}_{-1.0} \rightarrow [\textrm{S}/\textrm{H}]=0.33&^{+0.20}_{-0.27}
\end{split}
\end{equation*}
\noindent in which error calculations are similar to those explained in Appendix~\ref{app}.
\subsubsection{UV-soft SED}
Finally, we investigate how the photoionization solution changes when we employ the UV-Soft SED to the outflow. This SED is typically used for high-luminosity, radio-quiet quasars and is described in \cite{dun10,arav13}. Unlike the MF87 SED, which features a prominent UV bump, the UV-soft model excludes this feature, offering a cooler accretion disk with a power-law spectrum from the near-infrared to X-rays.

Figure~\ref{figUVsoft} shows the results of photoionization simulations of the outflow in 3C298 while considering the UV-soft SED and solar abundances. As this Figure shows, there is a single photoionization solution that can reproduce all observed column densities simultaneously: 
\begin{figure}
\includegraphics[width=\columnwidth]{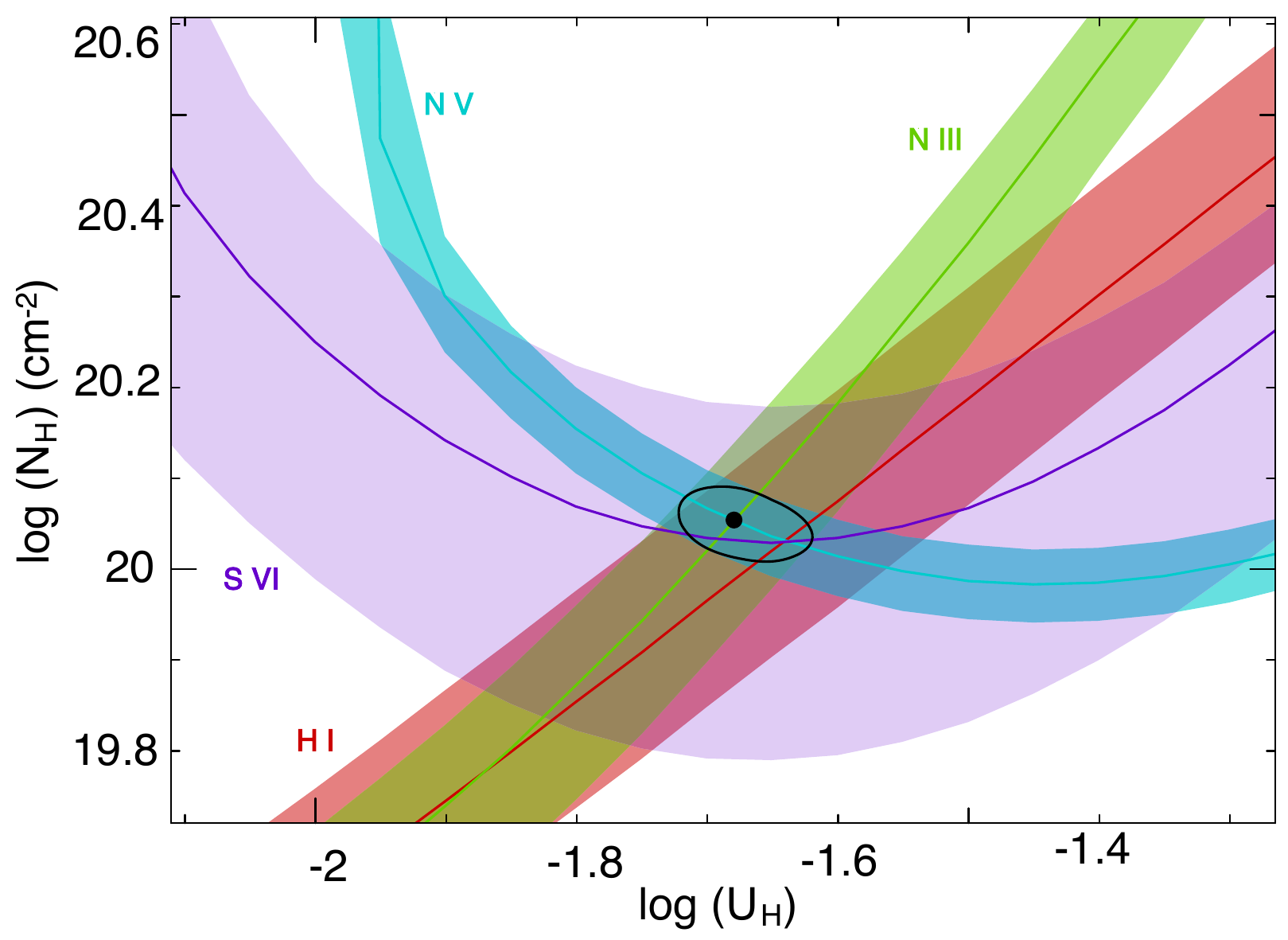}
      \caption{Same is Figure~\ref{figHSN}, utilizing the UV-Soft SED and solar abundances.}
         \label{figUVsoft}
\end{figure}
\begin{equation*}
\begin{split}
\frac{\textrm{Expected\  Abundances\ of\  N}}{\textrm{Solar Abundances of N}}(\textrm{with respect to H})=\\ 
\frac{1}{0.87 ^{+0.32}_{-0.22}} =1.15^{+0.40}_{-0.30} \rightarrow
[\textrm{N}/\textrm{H}]=0.06^{+0.13}_{-0.13}&
\end{split}
\end{equation*}
\noindent and:

\begin{equation*}
\begin{split}
\frac{\textrm{Expected\  Abundances\ of\  S}}{\textrm{Solar Abundances of S}}(\textrm{with respect to N}) = 0.93^{+0.4}_{-0.4} \\
\end{split}
\end{equation*}
\noindent resulting in:
\begin{equation*}
\begin{aligned}
\frac{\textrm{Expected\  Abundances\ of\  S}}{\textrm{Solar Abundances of S}}(\textrm{with respect to H})= \\
1.15^{+0.40}_{-0.30}\times 0.93^{+0.4}_{-0.4}=1.07^{+0.60}_{-0.54}
\rightarrow [\textrm{S}/\textrm{H}]=0.03&^{+0.20}_{-0.28}
\end{aligned}
\end{equation*}
To summarize the results derived from all three SEDs, we find that the nitrogen and sulfur abundances are consistent with solar values, with a small spread introduced by the choice of SED. Table~\ref{tab2} presents the final photoionization solution for each SED after adjusting nitrogen and sulfur abundances according to the calculations above. Based on this table, and by considering the variations introduced by the three SEDs, we report the total hydrogen column density of the outflow in 3C298 to be log$(N_{\textrm{H}}) = 20.05 \pm 0.3$ [cm$^{-2}]$. This value represents the median of the three measurements, with uncertainties chosen to encompass the full range of results from the different SEDs. Similarly, for the ionization parameter, we report log$(U_{\textrm{H}}) = -1.6^{+0.25}_{-0.15}$, where the uncertainties similarly account for the full variation across the SED models.

\begin{table}[ht!]
\caption{\label{tab2}Photoionization solution using three SEDs}
\centering
\begin{tabular}{lccc}
\hline\hline
SED&$\textrm{log}(N_{\textrm{H}}) [\textrm{cm}^{-2}]$&$\textrm{log}(U_{\textrm{H}})$&$\chi^{2}_{red}$\\
\hline
HE0238&$20.3^{+0.05}_{-0.06}$ & $-1.4^{+0.06}_{-0.05}$&0.01\\
\\
MF87& $19.8^{+0.04}_{-0.05}$ &$-1.7^{+0.06}_{-0.04}$&0.05\\
\\
UV-soft &$20.05^{+0.04}_{-0.04}$ &$-1.7^{+0.06}_{-0.04}$&0.02\\
\hline
\end{tabular}
\end{table}

\subsection{Electron Number Density}
\label{subsec:eden}
\begin{figure}
\includegraphics[width=\columnwidth]{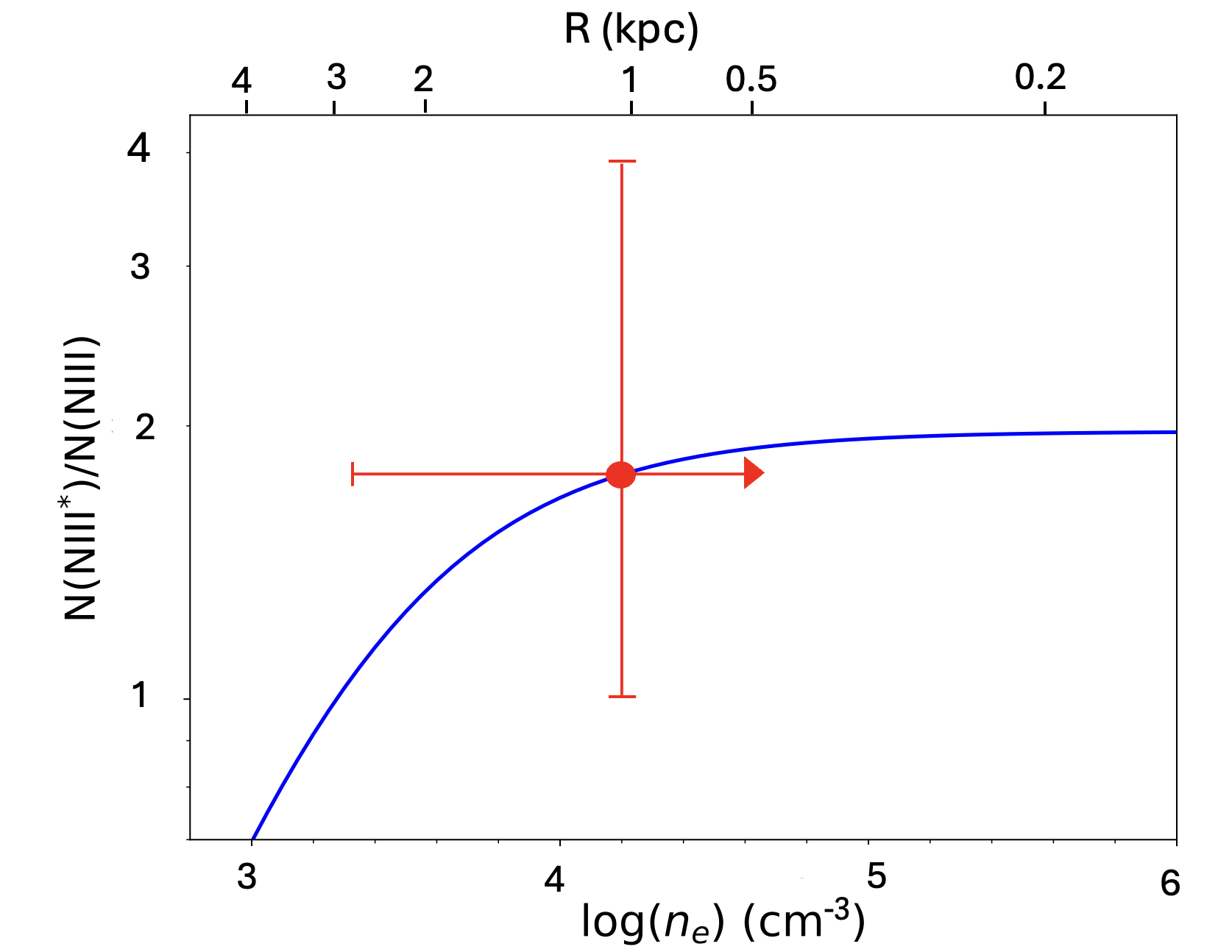}
      \caption{The blue curve shows the excited state (E=174.4 cm$^{-1}$) to resonance state column density ratio of \ion{N}{iii} vs. the electron number density from the Chianti atomic database for T=16000K. The measured ratio is shown by the red dot and the accompanying error bars by the vertical red lines. Note that the The 174.4 cm$^{-1}$ transition in \ion{N}{iii} arises from the $^2P_{3/2} \, \rightarrow \, ^2P_{1/2}$  transition within this ion.}
         \label{figchi}
\end{figure}

We use the Chianti 9.0.1 atomic database \citep{dere97, dere19} to calculate the abundance ratios of the excited state to the resonance state for \ion{N}{iii} as a function of electron number density (n$_{e}$). The Chianti atomic database uses the atomic data for \ion{N}{iii}, including energy levels, radiative transition rates, and collisional excitation rates. Then for a given temperature, Chianti calculates the column density ratio of \ion{N}{iii*} to \ion{N}{iii} as a function of n$_{e}$ (see blue curve in Figure~\ref{figchi}).
These calculations were performed considering a temperature of 16,000 K, which is predicted by Cloudy for the outflow in 3C298and by adopting the UV-soft SED. Figure~\ref{figchi} presents the results. Based on Table~\ref{tab1}, the column density of the excited state to that of the ground state is $\frac{N_{\text{\ion{N}{iii*}}}}{N_{\text{\ion{N}{iii}}}}  \approx 1.8$, (marked by a red dot in Figure~\ref{figchi}) corresponding to $\text{log}(n_{e})=4.2$ [cm$^{-3}$].  

To determine the uncertainties on the electron number density, we first determined the uncertainties in the column density ratio to be $\frac{N_{\text{\ion{N}{iii*}}}}{N_{\text{\ion{N}{iii}}}} = 1.8^{+1.9}_{-0.7}$, as outlined in detail in appendix \ref{app2}. These uncertainties, represented by vertical red lines in Figure~\ref{figchi}, were then used to calculate the corresponding errors in the electron number density: The upper uncertainty in the column density allows an $n_{\textrm{e}}$ of infinity, which is why the result is a lower limit on $n_{\textrm{e}}$.  The lower uncertainty allows a value of $\text{log}(n_{e})$=3.3 [cm$^{-3}$], as this is the point where the error intersects the theoretical curve.
Combining these results and based on the Chianti plot shown in Figure~\ref{figchi}, we conclude that $\text{log}(n_{e})\geq 3.3$ [cm$^{-3}$].

The top horizontal axis of Figure~\ref{figchi} shows the distance between the outflow  and the central source in parsecs, derived from the definition of the ionization parameter \citep{oster06}:
\begin{equation}
U_\textrm{H}\equiv\frac{Q_{\textrm{H}}}{4\pi R^{2}c~n_{\textrm{H}}} \Rightarrow R=\sqrt{\frac{Q_{\textrm{H}}}{4\pi c~n_{\textrm{H}} U_{\textrm{H}}}} \label{eq1}
\end{equation}

\noindent in which $Q_{\textrm{H}}$ is the number of hydrogen-ionizing photons emitted by the central object per second, $R$ is the distance between the outflow
and the central source, $n_{\textrm{H}}$ is the hydrogen number density, and $c$
is the speed of light. For a highly ionized plasma, we can assume $n_{\textrm{e}}\approx1.2~n_{\textrm{H}}$ \citep
{oster06}. For 3C~298, $Q_{\textrm{H}}=1.2\times 10^{57}$ s$^{-1}$, which is calculated using quasar's redshift, our adopted cosmology (see Section~1), and scaling the UV-soft SED to match the continuum flux of 3C298 at an observed wavelength of $\lambda=2650$\AA.

Using the ionization parameter resulting from the photoionization modeling with solar abundances (log$(U_{\textrm{H}})=-1.6$) and an electron number density resulted from Chianti database (log$(n_{\textrm{e}})=4.2\  [\textrm{cm}^{-3}]$), we locate the outflow at a distance of approximately 1050 pc from the central source. By considering the lower uncertainties on the electron number density, the outflow  could be located as far as 
$R_\textrm{{max}}=2800$ pc.

\section{Summary}
In this paper, we use archival data from the Hubble Space Telescope and employ photoionization modeling to determine the absolute abundances of nitrogen and sulfur in the absorption outflow seen in quasar 3C298. Our main conclusions can be summarized as:

\begin{enumerate}
     \item We employed three different SEDs to determine the chemical abundances in the outflow. Utilizing UV-soft and MF87 SED results in solar and super-solar abundances, respectively, which aligns with chemical abundances determined in other quasar outflows, as presented in Table~\ref{tab3}. However, by using HE0238 as the incident SED, our simulations indicate nitrogen and sulfur abundances are 60$\%$ of their solar values compared to hydrogen.  
     To come to a conclusion, each SED explored in this study produced a distinct regime of potential nitrogen and sulfur abundances. The three SEDs span an abundance range of 0.4-3 times solar. 

\begin{table*}
\caption{\label{tab3}Chemical abundances in outflows}
\begin{tabular}{lcccc}
\hline\hline
Object &[\textrm{N}/\textrm{H}]&[\textrm{O}/\textrm{H}]&[\textrm{C}/\textrm{H}]&[\textrm{S}/\textrm{H}]\\
\hline
Mrk 1044$^{a}$  &0.7&0.7&0.7&-\\
QSO J2233-606$^{b}$           & 1.1-1.3&0.5-0.9&0.5-0.9&-\\
Mrk 279$^{c}$            &0.4-0.66&$-$0.1-0.38&0.18-0.46&-\\
NGC 7469$^{d}$ &0.4-0.7&0-0.3&0.2-0.5&-\\
3C298 (HE0238)$^{e}$      &$-$0.36-$-$0.06&-&-&$-$0.36- $-$0.06\\
3C298 (MF87)$^{f}$ &0.09-0.38&-&-&$0.06-0.53$\\
3C298 (UV-soft)$^{g}$ &$-$0.07-0.19&-&-&$-$0.25-0.22\\
\hline
\end{tabular}
\tablefoot{This Table provides a comparison of abundance measurements, relative to solar values, drawn from multiple studies, including our present work. It is important to note that all abundance values are presented on a logarithmic scale.
\\$^{a}$ \cite{field05}
\\$^{b}$ \cite{gabel06}
\\$^{c}$ \cite{arav07}
\\$^{d}$ \cite{arav20}
\\$^{e, f, g}$ calculated in the current study. Adopted SEDs are indicated in parentheses.}
\end{table*}
    
    \item Performing photoionization modeling assuming three SEDs results in three pairs of solution for log$(N_{\textrm{H}})$ and log$(U_{\textrm{H}})$ (see Table~\ref{tab2}), or an average of log$(N_{\textrm{H}})=20.05^{+0.3}_{-0.3}$ and log$(U_{\textrm{H}})=-1.6^{+0.25}_{-0.15}$ for the outflow, in which we considered the systematic errors arising from various SEDs.
    
    \item We used the Chianti atomic database to derive the electron number density in the outflow. By comparing the ratio of the excited state to the resonance state of \ion{N}{iii}, we determine that $\text{log}(n_{e})= 4.2$ [cm$^{-3}$], which places the outflow at 1050 pc away from the AGN. However, by taking the uncertainties of the electron number density into account, we estimate that the electron number density is $\text{log}(n_{e})\geq 3.3$ [cm$^{-3}$]. This means that the outflow could be as far as 2800 pc away. One possible scenario to explain the sub-solar abundances results is that such outflow is located at this further distance and so less enriched than outflows closer to the center of the galaxy.
    
    \item Our findings suggest that the outflow system has a solar metallicity, with a 60$\%$ uncertainty. The 60$\%$ uncertainty in our metallicity determination arises from three different SEDs considered in this study—MF87, UV-soft, and HE0238. These SEDs represent different assumptions about the ionizing radiation field, which in turn influences the ionization parameter and column density calculations.
    
    \item \textcolor{magenta}{While sub-solar metallicities have been suggested in previous studies (e.g., \cite{arav07},\cite{pun18},\cite{marz23} and \cite{flo24}), our analysis contributes to refining the metallicity range by testing multiple SEDs.Our study serves as a complementary analysis that extends the findings of previous AGN feedback studies by exploring the effects of different SEDs on chemical abundance determinations.}

    \item For a given ionic column density of heavy elements, higher abundances result in a lower total hydrogen column density $(N_{\textrm{H}}$) required to reproduce the observed ionic column densities. Since $N_{\textrm{H}}$ is directly proportional to the kinetic luminosity in quasar outflows (equation 7 in \cite{borg12a}), a smaller $N_{\textrm{H}}$ leads to a reduction in the kinetic luminosity. This inverse relationship between abundance and kinetic luminosity indicates that adopting lower abundances could have significant implications for our understanding of galaxy evolution and feedback mechanisms.  
   \end{enumerate}

\begin{acknowledgements}
We want to express our sincere gratitude to the anonymous referee for their detailed and constructive feedback on our manuscript. We acknowledge support
from NSF grant AST 2106249, as well as NASA STScI grants AR-
15786, AR-16600, AR-16601, and HST-AR-17556. 
\end{acknowledgements}

%
%

\onecolumn
\begin{appendix}
\section{Error determination in the abundances calculations:}
\label{app}

The black dot in Figure~\ref{figHSN} suggests that in order to have a photoionization solution, \ion{H}{i} column density has to be $\log(N_{\text{Expected HI}}) = 16.10^{+0.08}_{-0.11}$. This is the value of \ion{H}{i} where the black dot is located and the errors are calculated based on the allowed uncertainties on the solution (shown by yellow crossed in Figure~\ref{figHSN}). By adopting the observed values ($\log(N_{\text{Observed HI}})$) and its uncertainties from Table~\ref{tab1}:

\begin{equation*}
\begin{aligned}
\log(N_{\text{Expected HI}}) = 16.10, \quad 
\log(\text{Upper uncertainty}) = +0.08, \quad 
\log(\text{Lower uncertainty}) = -0.11 \\
\log(N_{\text{Observed HI}}) = 16.32, \quad 
\log(\text{Upper uncertainty}) = +0.12, \quad 
\log(\text{Lower uncertainty}) = -0.11
\end{aligned}
\end{equation*}

\noindent By converting these values to linear scale, we have:
\begin{align*}
N_{\text{Expected HI}} &= 10^{16.10} = 1.26 \times 10^{16} \\
\text{Upper value (Expected HI)} &= 10^{16.10 + 0.08} = 1.5 \times 10^{16} \\
\text{Lower value (Expected HI)} &= 10^{16.10 - 0.11} = 9.8 \times 10^{15}
\end{align*}
\begin{align*}
N_{\text{Observed HI}} &= 10^{16.32} = 2.1 \times 10^{16} \\
\text{Upper value (Observed HI)} &= 10^{16.32 + 0.12} = 2.75 \times 10^{16} \\
\text{Lower value (Observed HI)} &= 10^{16.32 - 0.11} = 1.62 \times 10^{16}
\end{align*}

\noindent The discrepancy between the observed and expected  values can be determined as below:
\begin{equation*}
\text{Column density ratio} = \frac{N_{\text{Observed HI}}}{N_{\text{Expected HI}}} = \frac{2.1 \times 10^{16}}{1.26 \times 10^{16}} = 1.66
\end{equation*}

\noindent We can then calculate the relative uncertainties for the observed value of \ion{H}{I} column density as:

\begin{align*}
\text{Relative upper uncertainty (Observed HI)} &= 
\frac{\text{Upper value (Observed HI)} - N_{\text{Observed HI}}}{N_{\text{Observed HI}}} \\
&= \frac{2.75 \times 10^{16} - 2.1 \times 10^{16}}{2.1 \times 10^{16}} = 0.31
\end{align*}
\begin{align*}
\text{Relative lower uncertainty (Observed HI)} &= 
\frac{N_{\text{Observed HI}} - \text{Lower value (Observed HI)}}{N_{\text{Observed HI}}} \\
&= \frac{2.1 \times 10^{16} - 1.62 \times 10^{16}}{2.1 \times 10^{16}} = 0.23
\end{align*}

\noindent Similarly, we calculate the relative uncertainties for the expected value of the \ion{H}{I} column density:
\begin{align*}
\text{Relative upper uncertainty (Expected HI)} &= 
\frac{\text{Upper value (Expected HI)} - N_{\text{Expected HI}}}{N_{\text{Expected HI}}} \\
&= \frac{1.50 \times 10^{16} - 1.26 \times 10^{16}}{1.26 \times 10^{16}} = 0.19
\end{align*}

\begin{align*}
\text{Relative lower uncertainty (Expected HI)} &= 
\frac{N_{\text{Expected HI}} - \text{Lower value (Expected HI)}}{N_{\text{Expected HI}}} \\
&= \frac{1.26 \times 10^{16} - 9.8 \times 10^{15}}{1.26 \times 10^{16}} = 0.22
\end{align*}

\noindent The final relative uncertainties result from quadratically adding the relative uncertainties:

\noindent \text{Relative upper uncertainty (For the column density ratio)}=
\begin{align*} 
\sqrt{(\text{Relative upper uncertainty (Observed HI)})^2 + 
(\text{Relative lower uncertainty (Expected HI)})^2} \\
= \sqrt{(0.31)^2 + (0.22)^2} = 0.37
\end{align*}

\text{Relative lower uncertainty (For the column density ratio)} = 
\begin{align*}
\sqrt{(\text{Relative lower uncertainty (Observed HI)})^2 + 
(\text{Relative upper uncertainty (Expected HI)})^2} \\
= \sqrt{(0.23)^2 + (0.19)^2} = 0.30
\end{align*}

\noindent And finally, to calculate the absolute uncertainties in the column density ratio, we multiply the relative uncertainties by the ratio itself:
\begin{align*}
\text{Upper absolute uncertainty in the column density ratio} =\\ \text{Relative upper uncertainty (For the column density ratio)} \times \text{Column density ratio} \\
= 0.37 \times 1.66 = 0.61
\end{align*}
\begin{align*}
\text{Lower absolute uncertainty in the column density ratio} = \\\text{Relative lower uncertainty (For the column density ratio)} \times \text{Column density ratio} \\
= 0.30 \times 1.66 = 0.50
\end{align*}

\noindent Which results in the final value of:
\begin{equation*}
\frac{N_{\text{Observed HI}}}{N_{\text{Expected HI}}}= 1.66^{+0.61}_{-0.50}
\end{equation*}
\section{Error Calculations for the electron number density}
\label{app2}

Table~\ref{tab1} presents the measurements and uncertainties for \ion{N}{iii} and \ion{N}{iii*} as:
\begin{align*}
\log N_{\text{\ion{N}{iii}}} &= 14.41, \quad 
\log(\Delta N_{\text{upper, \ion{N}{iii}}})= 0.17, \quad  
\log(\Delta N_{\text{lower, \ion{N}{iii}}})= -0.30, \\
\log N_{\text{\ion{N}{iii*}}} &= 14.66,\quad 
\log(\Delta N_{\text{upper, \ion{N}{iii*}}})= 0.12,\quad 
\log(\Delta N_{\text{lower, \ion{N}{iii*}}})= -0.16,
\end{align*}

\noindent Using the values above, one can simply calculate the ratio of excited to resonance states for \ion{N}{iii} as:
\[
\log(\frac{N_{\ion{N}{iii*}}}{N_{\ion{N}{iii}}}) = \log N_{\text{\ion{N}{iii*}}} - \log N_{\text{\ion{N}{iii}}}=0.25 => \frac{N_{\ion{N}{iii*}}}{N_{\ion{N}{iii}}}= 1.8
\]

\noindent Using the logarithmic values, we can determine the uncertainties for the ratio of $\frac{N_{\ion{N}{iii*}}}{N_{\ion{N}{iii}}}$  as follows: 
\begin{align*}
\log(\Delta N_{\text{upper, ratio}}) &= \sqrt{(\log(\Delta N_{\text{upper, \ion{N}{iii*}}}))^2 + (\log(\Delta N_{\text{lower, \ion{N}{iii}}}))^2}= \sqrt{(0.12)^2 + (-0.30)^2} = 0.32, \\
\log(\Delta N_{\text{lower, ratio}}) &= \sqrt{(\log(\Delta N_{\text{lower, \ion{N}{iii*}}}))^2 + (\log(\Delta N_{\text{upper, \ion{N}{iii}}}))^2}= \sqrt{(-0.17)^2 + (0.18)^2} = 0.23.
\end{align*}

\noindent which can be converted to the linear scale:
\[
\text{Upper uncertainty} = 10^{\log(\text{Ratio}) + \log(\Delta N_{\text{upper, ratio}})} - 10^{\log(\text{Ratio})} = +1.9
\]
\[
\text{Lower uncertainty} = 10^{\log(\text{Ratio})} - 10^{\log(\text{Ratio}) - \log(\Delta N_{\text{lower, ratio}})} = -0.7
\]

\noindent  Therefore, the final results are:

\begin{equation*}
\frac{N_{\text{\ion{N}{iii*}}}}{N_{\text{\ion{N}{iii}}}}  = 1.8^{+1.9}_{-0.7}
\end{equation*}

\end{appendix}
\end{document}